\documentstyle[preprint,aps]{revtex}

\begin{document}

\begin{title}
{Surface States, Surface Metal-Insulator, and Surface Insulator-Metal
Transitions\thanks{Paper to appear on ''Electronic Surface
and Interface States on Metallic Systems'', ed E. Bertel and M. Donath
(World Scientific, Singapore, 1995)}}
\end{title}

\author{E. TOSATTI}
\address{International School for Advanced Studies (SISSA), \\
via Beirut 4, 34014 Trieste,\\
and \\
International Centre for Theoretical Physics (ICTP), \\
P.O.Box 586, Miramare, 34014 Trieste, Italy.}

\maketitle

\begin{abstract}
I present an informal discussion of various cases where two-dimensional
surface metal-insulator structural and charge-density-wave instabilities
driven by partly filled surface states have been advocated. These include
reconstructions of clean semiconductor surfaces and of W(100) and Mo(100),
as well as anomalies on the hydrogen-covered surfaces H/W(110) and
H/Mo(110), and possibly alkali-covered surfaces such as K/Cu(111). In
addition I will also discuss the opposite type of phenomena, namely surface
insulator-metal transitions, which can be argued to occur on $\alpha -$%
Ga(001), high-temperature Ge(111), and probably Be(0001).
\end{abstract}

\narrowtext

\section{\ Introduction}

Mainly because they historically involved different people and different
techniques, but probably also because each field had developed its own {\it %
ad hoc} theory, the subjects of surface structure and structural
transitions, surface electronic states, and surface lattice vibrations are
generally treated as disconnected.

As I see it, the main scope of this book, and of the hearty Bad Honnef
discussions which the book summarizes, is precisely to remind all of us of
how much these areas are, on the contrary, very much connected and
intimately tangled with one another. I will discuss here, more specifically,
two different instances, pertinent to crystal surfaces, where this tangling
becomes of dramatic importance.

The first instance, is that whenever there are partly filled electronic
surface states on a solid surface, then this sort of 2D metallic situation
may be unstable. The surface structure develops a tendency to rearrange --
in most cases it will reconstruct, giving rise to a new surface periodicity
--\ in such a way as to become a 2D insulator. My shorthand reference to
this class of systems will be to speak of a Surface Metal-Insulator (SMI)
transition. After general remarks, and qualifications on charge-density
waves, I shall devote Section 2 to a bird's eye view over some of the most
popular SMI cases. From these specific examples I shall also try to draw
more general wisdom, in some cases related to other contributions in the
present volume.

The second instance, is that there are now also examples (identified at
least in theory), of what may be seen, in an imprecise sense, as the
opposite phenomenon to SMI taking place. In systems where the bulk balance
between metal and insulator (or semimetal, or very covalent metal) is only
weakly tilted in favor of the latter, the balance will generally change at
the surface. Here, the fully metallic state has a chance to prevail, even
if only locally. Interestingly, in the cases where it does prevail, at
least at low temperatures, the final metallic coating at the surface of the
bulk nonmetal can be argued to have a function which is related to surface
states, in this case to non -existing ones. The function is in fact that of
removing the 2D metallic Fermi surface of the half-filled surface states
which the nonmetal surface would otherwise inevitably have if it did not
transform into a metal. In this sense, the surface metallization transition
is also driven by half-filled surface states, although in reverse. I will
call these cases Surface Insulator-Metal (SIM) transitions, to be discussed
in Section 3.

In the spirit of a series of introductive remarks to the more technical
subject matter of this book, this paper will be quite informal,and totally
devoid of either rigor or completeness. The accent will rather fall on a
variety of controversial points, in an attempt to stimulate progress,
whenever possible, through discussion.

\section{Surface Metal-Insulator (SMI) Transitions}

\subsection{General remarks}

It is obvious that a crystal surface, or more precisely a semi-infinite 3D
crystal, cannot generally be treated as simply a 2D system. However, it is
also true that atoms at a surface possess a lower coordination, sort of
half-way between 3D and 2D, which automatically requires a reassessment of
chemical bonding at a surface relative to that of a bulk. Furthermore in
some cases, like that of narrow well-defined surface electron and phonon
states, it is possible to map a surface onto a straight 2D problem.

If one glances at the periodic table of the (3D!) elements, one finds that a
majority of them are metals, under ordinary conditions. If instead of
three-dimensional, however, our world was only two- or one-dimensional, one
of the ways in which it would look different would be a dramatic decrease of
elemental metals. In lower dimensions, lower atomic coordination reduces the
bandwidth and therefore the electron kinetic energy gain which stabilizes
metals, while both electron-electron and electron-lattice interactions have
no smaller role, favoring an insulating state.

Electron-electron repulsion generally leads to magnetism, Mott insulators,
and the like. While there is no reason of principle why that should not be
relevant to surfaces, I am not aware well-established cases of surface
magnetic instability or reconstruction, at least for bulk nonmagnetic
sp-bonded elements. I will thus not discuss this possibility any further.

Apart from the additional exciting possibility of surface superconductivity
in a metal whose bulk is normal, of which again I do not know a proven case
, electron-lattice coupling tends to lead to a surface insulating state too,
of the regular band type. Starting from an ideal surface, the SMI transition
generally coincides with a structural, {\it reconstruction}, phase
transition. In analogy with other structural phase transitions, the SMI
transition may be gentle, or as one says {\it displacive}, or it may be
dramatic, or {\it reconstructive}. The former type involves a small
periodic distortion of the lattice at and near the surface, similar to a
frozen-in surface phonon, with a corresponding gapping of the electronic
surface states. The latter implies instead some kind of drastic
rearrangement of atomic populations and positions, including creation of
vacancies, adatoms, missing rows, etc., and a variety of rebonding
possibilities, with the only common end result of a large insulating gap
between filled and empty surface states.

The displacive SMI's can be generally classified as charge-density waves
(CDW's)\cite{overhauser2}, to be discussed below. The reconstructive SMI's
are less universal, more system-specific in their mechanism. I will briefly
allude, very incompletely, to some of them in the subsection devoted to
semiconductor surfaces.

\subsection{Charge-density waves : weak and strong-coupling}

Consider a surfaces which does, in its ideal{\it \ }bulk-like geometry,
possess well-defined partly filled electronic surface states, forming a 2D
Fermi sea. The density response function of this 2D Fermi sea is peaked at
the spanning, or nesting, vector 2$k_F$. The peak may be strongly enhanced
by two separate factors. The first, is the presence of large portions of the
Fermi surface, separated side by side by the same 2$k_F$ vector (nesting\cite
{lomer}). The second, is a large electron-electron exchange interaction\cite
{overhauser1}\cite{kohn68}\cite{halperin}. When the density response peak at
2$k_F$ diverges, this signals a CDW. The 2D Fermi surface response
influences also surface vibrations, so that Kohn anomalies are also
generated at momentum 2$k_F$. When the CDW sets in, the Kohn anomaly is
automatically large enough to drive a soft surface phonon, and there will
be a displacive change of structure. The surface lattice undergoes a
periodic distortion, of spatial periodicity close to $2\pi /$2$k_F$, and
amplitude proportional to the CDW order parameter $\Delta$. The outcome of
the CDW and its periodic lattice distortion is to eliminate as much of the
original 2D Fermi surface as possible, tending to pair electrons into new
completely filled states, separated by a band gap $2\Delta $ from new
completely empty states. Due to the lattice involvement, a large
electron-phonon coupling at 2$k_F$ will clearly favor CDW's, besides the
purely electronic factors such as nesting, and exchange.

The general ideas behind this kind of displacive SMI transitions (under
various names: excitonic insulator\cite{kohn68}\cite{halperin},
charge-density waves\cite{overhauser2}, band Jahn-Teller effect\cite{salem}%
\cite{englman}, Peierls instability\cite{peierls}, etc) go back to the
fifties and sixties. The possibility of their occurrence on surfaces was
raised in the seventies\cite{1974}\cite{kyoto}\cite{festkorper}\cite{karpacz}%
, and need not be discussed in great detail here. However, in view of the
quite common confusion and controversy, often purely semantic, which these
different terms and names appear to generate, it may be useful to remind
ourselves about CDW's in general.

At the general level, before entering a discussion specific to surfaces, an
important distinction must be made, between {\it weak-coupling CDW's},
(WCDW) which include Overhauser's original state\cite{overhauser2}, and {\it %
strong-coupling CDW's}, (SCDW), far more common in nature. In both WCDW and
SCDW the energy gain is electronic: the gap order parameter $2\Delta $
implies a reduction of electron kinetic energy, which in 1D is of order $%
\Delta ^{2\text{ }}\left| \ln \Delta \right| $ in the former, and of order $%
\Delta $ (asymptotically) in the latter\cite{karpacz}. The size of the
electronic gap relative to the phonon energies is the dividing line between
the two.

The small gap of WCDW's generally make them nonadiabatic phenomena, while
SCDW's can be essentially adiabatic, due to their large gap. Due to
nonadiabaticity, WCDW's ought to be sensitive to{\it \ isotopic substitution}%
, a test which, as far as I know, has never been tried. What drives thermal
disordering, and the eventual phase transition to an undistorted state
above a critical temperature $T_c$, is also different in the two cases. In
a WCDW, thermal disordering takes place largely due to electronic entropy.
When $k_BT\sim \Delta $, electron-hole pairs are excited across the gap,
which gradually loses its reason for stability, and disappears at $T_c$,
alongside with the periodic lattice distortion. The lattice plays a minor
role in disordering, it simply accompanies the electronic distortion. The
latter could in fact even take place, driven by exchange, in the absence of
a lattice distortion, as in spin-density-waves (SDW)\cite{overhauser1}.

SCDW's represent the opposite limit. Here, electron pairs are very strongly
bound. They constitute a kind of ordered superlattice of weak chemical
bonds, since here the relative size of the pair is comparable with the
lattice spacing, $\xi /a\sim \Delta /E_F$ $\sim 1$. There is a strong
tendency of the distortion to abandon the exact $2k_{F\text{ }}$periodicity,
in favor of some level of commensurability with the original lattice. The
electronic energy gain is entirely spread over the Brillouin Zone, (one
ought to keep in mind, however, that even in the WCDW,{\it \ the energy
gain does not strictly originate near} $k_{F\text{ }}$! The gain would only
be $\Delta ^{2\text{ }}$in that case, not $\Delta ^2$ $\left| \ln \Delta
\right| $ as it actually is. The $\left| \ln \Delta \right| $ factor is due
to electrons in the rest of the BZ, away from $k_{F\text{ }}$). In strong
coupling, the electronic gap is always much larger than $k_BT$, electron
entropy is irrelevant, and {\it lattice} entropy is instead what drives the
order-disorder transition\cite{inglesfield79}. At very high temperature, in
fact, the undistorted lattice is favored, as its higher symmetry implies a
larger entropy. At the critical point $T_c$, electron pairing is not
undone, as was the case in the WCDW. Pairs remain bound, but simply become
incoherent with one another. In other words, the lattice remains locally
distorted most of the time. However, the distortion fluctuates in space and
time, and true long-range order is lost.

It must be stressed that only at some higher temperature $T_o$, way above $%
T_c$, will the short-range distortion gradually dissolve, and a final
crossover take place to a truly undistorted, ordered state. Above this
crossover, fast thermal lattice fluctuations restore an ungapped band
structure, through a process which has not been studied in sufficient
detail, but which must be akin to {\it motional narrowing}\cite{kittel}.

Why is the above clarification useful for surface SMI practitioners?
Essentially, it should help us avoid unnecessary controversy, especially
between groups who emphasize different aspects, claiming to have altogether
different phenomena. For example band Jahn-Teller, or local bonding, which
are just other names for SCDWs, are sometimes quoted as plainly {\it %
alternative }to CDW's, while,as discussed, they do not represent
fundamentally different physics from WCDW's (except for adiabaticity, of
course). Similarly, surface order-disorder behavior close to a
deconstruction $T_c$ does not speak against a displacive mechanism. It is
the rule,on the contrary, close enough to $T_c$, for all continuous,
lattice-entropy driven structural phase transitions\cite{bruce}, which as
mentioned above, include SCDW's.

\subsection{ SMI transitions at semiconductor surfaces : mostly
reconstructive}

In surface physics, semiconductors are unique, in that essentially all of
their surfaces are reconstructed. This can be traced back to the strong
covalency of their bulk band structure. As is well-known, covalency requires
electronic states to hybridize, and then to overlap, so that filled states
correspond to bonds, empty states to antibonds, with a large energy gap in
between. At the surface, hybrid orbitals pointing outwards remain as {\it %
dangling bonds}, with energy at mid-gap, as appropriate for a collapse
between bonds and antibonds. The dangling bonds constitute therefore sharp
surface states (they fall in the bulk energy gap), and are only half-filled
(they are crossed by the Fermi level).

The uniqueness of covalent systems as opposed to, for example, regular
metals, is that bulk covalent bonding is so directional and inflexible, that
it will generally cost too much energy to locally readjust wavefunctions at
the surface, without some major lattice rearrangement. In other words, it is
not possible to fully {\it rehybridize} away the very costly half-filled
dangling bonds. Radical rehybridization in fact would cause too much
energetic disruption, particularly to the ''back bonds'' between first and
second atomic layer. Orbital readjustment is, on the contrary, generally
possible and very easy on the surface of a good metal, which has no bulk
energy gap, and where covalency is unimportant. Semiconductor surfaces are
thus stuck with half-filled dangling bonds, which constitute a nearly ideal
2D metallic system, prone to SMI instabilities of one kind or another. They
are so unstable, that the strict necessity for the surface to passivate
itself, and become a 2D insulator, is the key to understanding in essence
all microscopic semiconductor surface physics, including structure,
vibrations, and electronic states.

As a consequence, the tendency towards an SMI instability, and the
associated energy gain, are very large on semiconductor surfaces. A typical
order of magnitude is 0.2$\div $0.3 eV /surface atom, calculated for clean
(111) surfaces of diamond-structure group IV semiconductors. Due to that,
there seems to be on semiconductors, contrary to my original suggestions\cite
{1974}\cite{kyoto}\cite{festkorper}, relatively little room for weaker
displacive reconstructions such as WCDW's. I will return to this point,
however, at the end of this section.

The majority of stable semiconductor surfaces instead undergo large
reconstructive SMI-driven rearrangements. A well-known example is Pandey's $%
\pi $-bonded chain model,\cite{pandey} which explains the (2$\times $1)
state of the cleaved (111) surfaces of diamond\cite{C(111)}, Si\cite{Si2x1}
Ge\cite{takeuchi1}, and $\alpha $-Sn (111)\cite{Sn2x1}, which are free of
adatoms. Another example is that of the adatom reconstructions, with the DAS
model by Takayanagi{\it \ et al}\cite{DAS} for Si(111)7$\times $7, and the
simpler but related adatom model for Ge(111) c(2$\times $8)\cite{c(2x8)}. In
the (2$\times $1) $\pi $-bonded chain model, one back bond per surface cell
is broken and shifted, so that a good resonant overlap of dangling bonds
becomes possible along polyacetylene-like surface chains. On diamond (111),
there appears even to be a weak dimerization of the chain\cite{C(111)},
which is then totally similar to polyacetylene\cite{ssh}. In adatom
reconstructions instead, strikingly, the reconstruction-induced real-space
electron pairing is {\it ionic}. Here, concerted and opposite relaxations (
inwards for adatoms, outwards for free first-layer atoms, called restatoms)
lead to transfer of an electron from the adatom, whose dangling bond
remains empty, to the restatom, whose dangling bond becomes doubly occupied.
This is an example of Anderson's ''negative U''\cite{anderson}, not the
only one in semiconductors and semiconductor surfaces\cite{anc1}\cite
{biarritz}. Finally, besides adatoms, also surface vacancies are predicted
to have an interesting interplay with reconstructions of semiconductors. One
example is the strong $\sqrt{3}\times \sqrt{3}$ buckling distortion which
has been predicted to decorate the neighborhood of a vacancy on ideal Ge(111)%
\cite{anc2}

Are there then any CDW's on semiconductor surfaces? Actually, the answer is
affirmative : but they are all very strong coupling SCDW's, so strong that
they hardly ever deconstruct at a reasonably low temperature. The dimer
reconstruction of Si and Ge(100)\cite{dimer}\cite{dimer1} can be seen as a
very large amplitude SCDW. The up-down buckling which stabilizes the $\pi $%
-bonded chain $(2\times 1)$ reconstruction of Si, Ge, and $\alpha -$Sn (111)%
\cite{Si2x1}\cite{takeuchi1}\cite{Sn2x1}, and the dimerization which
stabilizes that of diamond (111)\cite{C(111)} are also examples of SCDWs,
since the chain would be a 1D metal without them.

Finally, detailed calculations show that even on the ideal clean Ge(111)
surfaces a simple alternative SCDW-type state, which consists of a regular
up-down buckling (leading to a (2$\times $1) ionic reconstruction) does in
fact exist. It is an energy extremum,{\it \ }only slightly higher than
either the $\pi $-bonded chain or the adatom-restatom states. Hence the SCDW
is in this case only a local minimum, or perhaps a saddle point.
Interestingly though, even if it never makes it to the ground state, this
state may still not be completely irrelevant. As revealed by recent {\it %
ab-initio} simulations, it appears to play the role of a transition state in
the evolution between the ideal surface and the final $\pi $-bonded chain
state \cite{takeuchi1}.

\subsection{ W(001), Mo(001): good examples of SCDW's}

The bcc group VA and VIA metals are those, among all transition metals which
bear the closest resemblance to semiconductors. There is in their bulk
electronic structure a certain degree of covalent bonding involving the
d-states, which shows up clearly, even if indirectly, in the electronic
density of states\cite{papa}. It shows, as a function of energy, two rather
well separated groups of states. Between them, a region of relatively low
density of states, or pseudogap. The two large groups of electronic states
are, again, superpositions of directional d-state hybrids, of bonding and
antibonding character respectively\cite{harrison}. In the refractory metals,
Mo and W, the bonding states are totally filled, and the Fermi level falls
in the pseudogap, below the antibonding ones. This qualitatively explains
their high stability, hardness, and melting point. On the other hand, the
Fermi level cuts to some degree into the bonding bands in the case of V, Nb,
Ta (which have one electron less) and also in Cr, where s states are
comparably deeper than in Mo, W. The incomplete filling of the bonding
d-states is clearly connected with bulk anomalies, including the
antiferromagnetism of Cr, and the superconductivity of V, Nb, and Ta.

Amusingly, the behavior of the (001) surfaces of these bcc metals is just
reversed (at least for the nonmagnetic ones) with respect to their bulk. On
these faces, it turns out, there are dangling-bond-like surface states\cite
{weng}, analogous to the case of semiconductors, and here too, they fall at
mid (pseudo-) gap. Moreover, the ideal surface electronic structure is quite
similar for pairs of bcc transition metals which differ by just one
electron, like Ta vs. W, and Nb vs. Mo. Because of that one electron
difference, however, the surface states are partly filled in W and in
Mo(001), but totally empty in Ta and Nb(001)\cite{louie}. Electron
counting, the very reason for bulk instability of Nb and Ta, has now become
a reason for surface {\it stability}. Exactly the opposite is true {\it \ }%
for Mo and W{\it \ : }on the (001) surfaces of these very stable metals, one
now expects electronic instabilities, and surface phonon anomalies, possibly
leading to some kind of surface-state driven SMI transition. The
reconstructions which Estrup {\it et al.} \cite{felter}, and King {\it et al.%
}\cite{debe}{\it \ } found in the 70's have unmistakeably that origin, and
are due to those partly filled surface states, \cite{1978}\cite
{inglesfield78} as later confirmed by many experiments (rather well
summarized in Hulpke's paper\cite{hulpke95}) and also by microscopic
calculations.\cite{freeman}\cite{ho}\cite{krakauer}

Still, to this date, lingering questions seem to have remained around. Are
these actually CDW's or are they not? For example, is the reconstruction
periodicity (exactly c(2$\times $2) for W(001), and close but slightly
different from c(2$\times $2) for Mo(001)) dictated by the 2D Fermi surface,
as originally proposed\cite{1978}and as appropriate for WCDW's\cite{ho}, or
is it more related to coupling with the underlying bulk lattice and its
surface phonons, as expected in a SCDW?\cite{1980}\cite{1987}. Is the
finding of critical fluctuations at an apparently {\it incommensurate}
wavevector, seen by elastic He-scattering on W(001) near (1/2,1/2) $\pi /a$
just above $T_c$, an evidence for 2D Fermi surface effects\cite{ernst87},
or could it again be a purely lattice effect? Finally, a basic question
sometimes raised is, is the deconstruction of W(001) order-disorder like, or
order-order like\cite{king83}\cite{ernst92}?

I believe it has been made very clear during the past decade, that the SCDW
scenario is the correct one. The key role of the half-filled surface states
is unquestionable, as shown in particular by the absence of anomalies for
Nb(001)\cite{melmed}\cite{dewette}, in agreement with predictions \cite{1978}%
,and by their disappearance from Mo and W(001) upon H adsorption \cite
{ernst92}. A large phenomenology (also in agreement with later electronic
structure calculations) suggests in addition that coupling is relatively
large, and the electron pair correlations length is correspondingly short.
Under these circumstances, the effect of electronic forces may be mimicked
in a first approximation by an additional short-range force between surface
atoms. In other words, the net effect of half-filled states onto ionic
motion can be cast into a pure, unretarded classical potential, much in the
same way that an electronic bond in a solid or a molecule can often be
replaced by a classical spring\cite{1980}\cite{1987}. If this picture is
admittedly crude, it not outrageously so in strong coupling, and has the
enormous advantage to allow an accurate treatment of everything else,
related specifically with the position, dynamics, and statistical
mechanics of the ions. In fact, for everything except electronic structure,
which obviously it does not treat, it is found that this model works
quantitatively, correctly describing how the distortions is largely, but not
entirely in the first layer\cite{1987}, the surface phonons, including a
very elusive mode at k=0\cite{erskine}\cite{wang87}, the thermal
deconstruction behavior, which exhibits both displacive and order-disorder
features\cite{europhys88}, the soft phonon behavior with temperature\cite
{prb88}\cite{fas88}\cite{han}\cite{ernst92}. It also predicts a large
difference between a deconstruction $T_c$ around 280 K, and a much higher
crossover $T_o$, expected around 800 K on W(001)\cite{unpubl91}.

For Mo(001), the classical force model suggests that lack of exact c(2$%
\times $2) commensurability (formerly described as incommensurate\cite
{felter}, and subsequently found to be higher order c($7\sqrt{2}\times \sqrt{%
2}$)R45$^o$ commensurate\cite{smilgies91}\cite{hildner}) need not be
strictly due to a Fermi surface surface effect\cite{1978}, but may also
arise due to competition between in-plane and bulk-mediated purely lattice
forces, giving rise to soft phonon modes away from, but close to,
high-symmetry k-points \cite{1980}\cite{M5}\cite{Mo88}. It is in fact
possible, by modifying slightly the interaction parameters \cite{wang95}
from the somewhat arbitrary values adopted in an earlier study \cite{Mo88},
to reproduce rather closely the c($7\sqrt{2}\times \sqrt{2}$)R45$^o$%
structure recently reported for Mo(001)\cite{robinson}. Very reasonable
geometries are also obtained \cite{roelofs}, using the phenomenological
potentials developed by Carlsson\cite{carlsson}.

As for electronic states, their accurate description at T=0, and their
thermal evolution at and across the transition, the situation is less clear.
Local density calculations, at the best of their accuracy, confirm that a
displacive reconstruction lowers the energy, and suggest a rather large
gain in the order of 0.1 eV per surface atom at T = 0\cite{krakauer} (
against the 0.03 eV which realistically explain the observed transition
temperature\cite{europhys88}\cite{prb88} and distortion magnitudes\cite{1987}%
). Photoelectron spectroscopy has been contradictory, with papers first
suggesting the irrelevance \cite{campuzano}, and more recently the relevance
of the 2D Fermi surface and its gapping across $T_c$ both on W(001)\cite
{kevan90}, and Mo(001)\cite{chung}.

In this connection, I would like to call attention to the fact, often
overlooked (but not always: see Ernst {\it et al.}\cite{ernst92}), that the
nature of the SCDW phase transition, even if displacive, implies a
disorder-order crossover regime between $T_c$ and{\it \ }$T_o$, with {\it \
}$T_o\gg T_c$. Thermal evolution of the electronic states should begin with
sharp, gapped surface bands well below $T_c$ and end again with relatively
sharp ungapped bands, {\it not above }$T_c${\it, but above }$T_o$ ! Only
above {\it \ }$T_o$ ($\approx $800 K for W(001)), does in fact short-range
reconstructive order disappear, the motional narrowing of electronic bands
become complete, and the surface behave as truly undistorted. It is
therefore not a surprise that photoemission evidence should appear messy,
and confusing, anywhere between 100 K and 800 K for W(001), or, rescaling
down a factor of roughly two, between 50 and 400 K for Mo(001).

Nobody has unfortunately given a useful theory yet, or a simulation, of what
to expect of surface electronic states and their spectral function inside
the crossover temperature range. However, it is unlikely that experimental
results in this range should be amenable to a simple interpretation based on
a mean-field picture, which is what has been tried so far. On the other
hand, a mean field-like picture should work better when comparing data
between, say, 50 K and 1000K, since both temperatures should fall outside of
the crossover range. Should this comparison be possible in the future, we
would get a clearer message about the reconstruction of W(001) and Mo(001)
from electron spectroscopy.

Related comments apply to He-scattering results, discussed in great depth
by Hulpke and his collaborators for W(001)\cite{ernst92}\cite{hulpke95}.
Here, I would like to take up a relatively controversial aspect, namely the
apparently incommensurate He scattering peak just above{\it \ }$T_c$ \cite
{ernst87}\cite{ernst92}, which is invisible in LEED, and is attributed to
purely electronic incipient CDW scattering. Much as their overall viewpoint
is attractive, I would be considerably less certain about the actual 2D Femi
surface origin of their observed incommensurate scattering peak, than they
seem to be. In their support, it must be admitted that He scattering is
nearly uniquely and exquisitely sensitive to the Fermi surface (simply
measuring a particular surface electronic correlation function). The
evidence that Fermi level surface electron correlations becomes
incommensurate above $T_c$ must therefore be accepted as a fact. If other
surface tools miss that, it must be because they are less sensitive to
surface electrons near the Fermi level, not because the He results are
artifacts. However, it must also be considered that if surface electrons
appear incommensurate, that means something about the ions too, since
electrons are in the end not really free to do what they will. Surface
electrons are, adiabatically or not, tightly connected with surface ions,
which in turn we know to be, at these temperatures between 300 and 400 K,
barely deconstructed out of the c(2$\times $2) state, still strongly
short-range ordered, slowly fluctuating, and very much in the crossover
region. There is no such thing as a clean, free 2D Fermi surface, in this
regime.

My inclination is to believe, without a definite proof for the time being,
that the He-scattering incommensurability could in the end be traced back to
the surface ionic distortions. In very different surfaces, namely (1$\times
$2) reconstructed Au(110) and Pt(110), there is a very similar shift of the
half-order spot remnants, towards an apparent incommensurability, which
develops linearly just above $T_c$.\cite{lapuj}\cite{cvetko}\cite{vlieg}\cite
{kern}. The reason for that can be understood via SOS models\cite
{bernasconi93}, which show that due to proliferation above $T_c$ of an
excess of one type of steps (antiphase domain walls), say 1$\times $3
steps, over the opposite type, say 1$\times $1 steps, leads a shift of the
1/2 order spot towards smaller k-vector abobe deconstruction\cite{mazzeo}.
This feature is sometimes referred to as ''chirality''\cite{dennijs}.

Also on the c(2$\times $2) distorted surface of W(001), one can consider the
role of antiphase domain walls\cite{1987}. Chirality is expected here too,
because of the obvious inequivalence of ''heavy'' and ''light'' walls. For
example, the current model for the c($7\sqrt{2}\times \sqrt{2}$)R45$^o$
reconstruction of Mo(001), as determined from X-ray data\cite{robinson}
consists of a regular array of {\it strictly light walls} (in fact, one
every three zig-zag chains). Therefore, if domain wall proliferation
is, as is likely, an important process in surface deconstruction of W(001)
just above $T_c$, then the explanation of incommensurability could be, {\it %
mutatis mutandis}, just the same as on (1$\times $2) reconstructed Au(110)
and Pt(110). The main difference being that while on Au(110) the domain
walls are actual surface steps, on W(001) and Mo(001) they are phase slips
of a small in-plane periodic lattice distortion.

On the experimental side, if this speculation is correct, I would expect
that a surface scattering tool, sensitive to ions instead of electrons, but
still sensitive enough to short-range order effects, should be able to pick
up again the same incommensurability above $T_c$,as the He scattering probe
does.

Another point I would like to make is that it can be argued, interestingly,
that a wall will necessarily involve a local {\it vertical} displacement
too. This can be understood equivalently in terms of (a) the lattice
dynamics of the flat, unreconstructed surface, implying mixing of the
so-called M$_5$ (in-plane) and M$_1$(vertical) modes\cite{1980}\cite{M5}\cite
{1987}; (b) the soliton lattice model \cite{Mo88}, where the phase
quadrature between M$_5$ and M$_1$ modes implies a local M$_1$ distortion
creeping in between two antiphase M$_5$ regions; (c) local strain relief at
the wall\cite{1987}\cite{heine88}, which is the easiest to explain.
Basically, a surface atom in the center of an antiphase wall, has a
displacement whose in-plane component is zero by symmetry, but whose
vertical component can generally be finite. At a light wall, all four
neighboring surface atoms have moved {\it away}, creating an effective
potential trough, into which the center atom must fall. Hence, I expect a
{\it downward }displacement in the center of a light wall, and conversely an
{\it upward }displacement at a heavy wall center. If light walls
proliferate, and also move, they should carry with them this downwards
surface buckling.

It would be interesting to try to pick up these vertical displacements,
either in the reconstructed state of Mo(001), or on the just-deconstructed
state of W(001) slighly above $T_c.$ As it turns out, unfortunately, X-ray
data are relatively insensitive to vertical displacements, so much that the
amount of buckling on Mo(001) is unknown, at least so far\cite{robinson1}.
The large sensitivity of He scattering, conversely, may offer an additional
clue as to why a wall proliferation should be revealed so effectively in
that experiment.

On the theory side, this idea could in principle be pursued quantitatively
via simulations based on classical hamiltonians ({\it ab initio }ones being
far too time-consuming, at this stage). The required simulations would need
considerably larger sizes than those employed in our earlier studies \cite
{europhys88}\cite{prb88}. Some evidence for incommensurability, had in fact
occasionally been seen at short times in those earlier simulations, even
with their very limited cell sizes.\cite{wang thesis}.

A final question pertinent to that scenario is : what happens to the surface
state at an antiphase domain wall? The standard soliton picture\cite{ssh}
would suggest a mid-gap state, localized at the wall. In this case, it would
correspond more or less directly, to the dangling bonds of the center atom
in the wall. It would possess a dispersion parallel, but not perpendicular
to the wall. Moreover, the vertical displacement of the center atom would
control its energy, and therefore its electron filling. Altogether, it would
seem exciting to pursue this concept, and attempt a detection of such a
domain wall surface state.

\subsection{Clean Cu(111) : incipient CDW's on noble metals?}

The (111) face of noble metals : Cu, Ag, Au, possesses an s-like surface
state, with a good, roughly free-electron like, parabolic dispersion near
the $\Gamma $- point\cite{Cu(111)}.The state is partly filled, with a modest
2D electron density, corresponding to Fermi energies $E_{F\text{ }}$ of 0.44
, 0.12, and 0.5 eV for Cu, Ag and Au(111) respectively. The 2D Fermi surface
is correspondingly spanned by a rather small $2k_F$ ($\sim 0.4\AA ^{-1}$for
Cu(111)), very close in fact to the diameter of the famous (111) ''neck'' of
the 3D noble metal Fermi surface\cite{shoenberg}. Physical effects connected
with this his surface state are further addressed by Inglesfield and his
collaborators\cite{inglesfield95}

Probably due to a strong screening of exchange by the underlying bulk, to
the insufficient nesting which provided by the approximately circular shape,
and to depth mismatch, the 2D Fermi surface is unable to drive a CDW (or an
SDW) on these surfaces, which remain undistorted and metallic. The surface
phonons at $2k_F$ decay as $\exp (-z/z_{2k_F})$ below the surface plane (z =
0) with a penetration depth $z_{2k_F}$ which is large, since $2k_F$ is
small. The corresponding penetration depth of the surface electronic state
at $\pm k_F$ is not the same. On Cu(111) it is possibly smaller, in other
cases it might be larger. Quite generally, the {\it mismatch} between
penetration depths of surface electron and phonon states is a likely factor
to reduce the coupling between the two, and suppress or limit the occurrence
of surface CDW's. A relative paucity of CDW examples has been noted by Kevan%
\cite{kevan95}. Depth mismatch is likely to be one important factor behind
it.

On Cu(111), where there is no stable CDW, and moreover no strong Kohn
anomaly has been found in the surface phonon spectrum, there is nonetheless
a direct evidence that the surface response is high at $2k_F$. Recent
low-temperature STM topographs have shown spatial oscillations, with the
right periodicity $2\pi /2k_F$ $\approx 15\AA $, decorating the neighborhood
of Fe surface impurities.\cite{eigler}. In the light of these results, SDW's
or CDW's might perhaps be suspected to be incipient on this and similar
surfaces.

In reality, however, CDW's are probably still far from incipient. It should
be again stressed in this regard that STM and He scattering are two tools
which {\it \ greatly overemphasize the Fermi surface}, since they couple
directly to it. The underlying overall energetics associated with these
surface oscillations, including for example the resulting indirect
interaction between surface impurities via bulk electronic states\cite
{einstein}\cite{schrieffer}, and via surface states \cite{1976}\cite{kohn}%
,is in fact likely to be small in this case\cite{kevan95}. A well-defined
Kohn anomaly, i.e. a dip in the surface phonons at $2k_F$, remains a
mandatory piece of evidence, for CDW's incipiency.

\subsection{ Alkali-covered noble-metal surfaces :K/Cu(111)}

Hirschmugl {\it et al} have recently uncovered an unexpected anomaly in the
IR spectrum of the alkali-covered noble metal surface K/Cu(111)\cite
{hirschmugl}\cite{persson95}. They find, for K coverages sufficient for the
alkali to begin forming metal islands, $\theta _K$ $\geq 0.4$ monolayers in
this case, a sharp IR {\it antiabsorption }peak, at a photon energy of about
0.1 eV. The antiabsorption nature of the peak signifies, as shown by Persson%
\cite{persson}\cite{persson95} that the mode involves an in-plane charge
oscillation, of a type conventionally believed to be totally invisible in
IR. What is to be explained, is {\it what charge is oscillating}. At such a
high frequency, it cannot be a lattice mode, and the oscillation must
therefore be electronic. Formation of a CDW has been invoked, favored by the
dense metallized alkali layer. The CDW is pinned by some mechanism, such as
disorder or roughness, so that its hindered translation acquires a finite
frequency, of just 0.1 eV.

At first, the proposition looks attractive. After all, there was already a
surface state on clean Cu(111), with a possibly incipient CDW (see above),
before the alkali was deposited. Maybe the surface state can get a little
flatter, or it can get better coupled, or else maybe the metallic alkali
monolayer may screen out the electric field\cite{hirschmugl}, tilting the
balance a little towards CDW's. Maybe; but a few extra facts make it all a
little less simple, to say the least.

One fact is that the effect is absent for good quality surfaces, and shows
up only after a dose of scratching and crude roughening. Another fact is the
remarkable reproducibility of the narrow antiabsorption frequency, which
ought to show at least inhomogeneous broadening if due to CDW pinning. Yet
another, is that there is experimental proof that alkali coverage alters
the surface state picture very substantially.

The behavior of noble metal (111) surface states with alkali adsorption has
been recently quite well characterized\cite{fauster}\cite{fauster1}. For
very small coverage $\theta \leq 0.1$, the clean-surface state described in
Section 2.4 survives, although slightly pushed down in energy, closer to
bulk states. In this regime, one might perhaps imagine a tendency for the
alkali adatom density to become modulated in some interesting relationship
with the $2k_F$ charge oscillation they generate in the surface. As coverage
increases, however, the surface state is pushed right into the bulk
continuum, {\it and it disappears altogether}. At the same time, the work
function decreases, and an image state, initially several eV's above $E_F$,
is descending. At the metallization coverage, the alkali metal condenses
into islands of coverage one, and this state suddenly drops below the Fermi
level, forming a new partly filled surface state. It has about the same $%
2k_F $ of around 0.4 $\AA ^{-1}$ as for the clean surface, but with a mass
perhaps 4 times larger, whence $E_{F\text{ }}$ is correspondingly smaller,
0.11 eV instead of 0.44\cite{K/Cu(111)}. The character of this state is now
partly alkali, and partly Cu, and its penetration depth is at least a factor
2 smaller than the clean surface state.

This new surface state could now drive a CDW, and it could do that
effectively, since the factor 4 in the mass will also multiply the response
at $2k_F$. But why should this CDW require roughness, or disorder, or
possibly impurities, and refuse to show up on clean, smooth surfaces? And
why should an inhomogeneously pinned CDW yield nonetheless a single,
well-defined pinning frequency? Is there perhaps a reason why the new Fermi
energy and the antiabsorbtion peak frequency are so close? Shouldn't a
diffraction study pick up signs of at least some, even blurred, CDW
scattering, anyway, before we believe it?

To an outside spectator, the 0.1 eV feature actually looks more like a 2D
plasmon, localized and made accessible to optical detection by disorder and
imperfections. Being determined by the full coverage 2D electron density the
(wavevector integrated) plasma frequency is sharp, and independent of mean
coverage, disorder, etc. Is this a true alternative explanation?

The answer to these questions and speculations is unknown, and must be
postponed to the future.

\subsection{ One-dimensional Kohn anomalies on surfaces: H/W(110) and
H/Mo(110)}

If I must quote two surfaces where a textbook-clear fingerprint of
surface-state induced giant Kohn anomaly and/or CDW is clearly visible,
then those surfaces are to my mind H/Mo(110), and H/W(110).

The evidence, first found by Hulpke {\it et al.}\cite{hulpkeH}\cite{hulpke95}%
is of an extremely sharp, one-dimensional-like dip in the surface phonon
spectrum. Comparison with e.g., the phonons of the quasi- one-dimensional
chain compound KCP \cite{carneiro} is striking : it is hard to tell the two
apart. Similarly striking is a comparison with the calculated
phonon/phason/ampliton spectrum of a single 1D chain, carried out long ago
within mean-field theory, \cite{giuliani1}\cite{giuliani2}and reproduced in
Fig.\ref{fig:one}.
The extremely large slope $\left| d\omega /dk\right| $ of the lower surface
mode near the anomaly at {\bf k = Q }simply demands an electronic
explanation (a Fermi velocity is typically 10$\div $100 times larger than a
sound velocity), and disfavors at the outset the alternative (and otherwise
attractive) non-electronic interpretations for the anomaly\cite{balden}.A
substantial embarrassment to the surface state community comes from the fact
that there was, prior to the discovery of this anomaly, no compelling
reason, as to why of all surfaces, precisely H/W(110) should suddenly
develop, at full coverage and only there, such a strong Kohn anomaly, and a
{\it one-dimensional} one{\it \ }on top of that! There are in fact partly
filled surface states on this surface, and their 2D Fermi surface, as
extracted from photoemission data, had been previously reported\cite{kevan89}%
. But the nesting suggested by that data, was neither strong, nor did it
possess a k-vector strikingly close to the observed phonon anomaly.

Recent calculations on H/Mo(110) by the Berlin group\cite{kohler}are more
illuminating, showing a 2D Fermi surface with a nesting vector in better
agreement with the phonon anomaly, which is therefore likely to be indeed a
giant Kohn anomaly. These calculations suggest the following overall
picture. The partly filled surface electronic states and their 2D Fermi
surfaces are already present on clean Mo(110). There, however they are still
too close in energy to the bulk states, their penetration depth too large,
to be important for surface phonons. Upon H-adsorption, the H orbitals
hybridize and mix covalently with the surface states, and push them {\it down%
} in energy as a consequence, right into the middle of the surface gap for
projected bulk states. The surface state character is thus greatly enhanced,
the penetration depth now comparable with the lattice spacing, and the
nesting vector closer to the observed phonon anomaly, with fairly good
parallel portions. All in all, the calculation results come way closer to
explaining the phonon anomaly, than the photoemission-derived 2D Fermi
surface does.

Why should we trust a calculation more than a piece of data, when it comes
to arguing about the 2D Fermi surface? I believe we should indeed do that.
The key lies in the strong possibility, which I advance, that while the
surface used for the calculation is truly undistorted, that used for
experiment is only undistorted {\it on average}, must be very distorted,
instantaneously and locally. That this should be the case, follows from the
strong, slow fluctuations which necessarily accompany the soft phonon
branch, whose experimental energy takes a plunge from the 20 meV or so of
the Rayleigh wave, down to a dip of 1 or 2 meV only, at Q = 0.93 \AA $^{-1}$%
. Also the correlation length implied by this dip is very large, of order
100 \AA\ or so, and is probably terrace-size limited, as suggested by the
overall temperature-independence of the anomaly.

Now, if in actual fact the surface is distorted,{\it \ pieces of the 2D
Fermi surface originally connected by the distortion wavevector{\bf \ Q}
must disappear} due to Bragg scattering. Their remnants are modified, and
generally shifted elsewhere in k-space. Hence, the apparent 2D Fermi surface
in presence of a distortion, even if local and fluctuating, differs
maximally from the ideal one precisely where nesting k-vectors are closest
to that of the local distortion periodicity. This means, that precisely
due to the anomaly, we should {\it not} expect to see apparent nesting for
{\bf k = Q, }which is {\bf \ }what photoemission finds! Future
calculations, experiments, and simulations, should be able to enlighten
further this aspect, which seems crucial in order to make contact with
photoemission.

What about hydrogen? Do the adsorbed hydrogens play any additional role,
other than just conveniently shifting pre-existing surface states so that
they yield an anomaly? EELS data suggest an affirmative answer, with an
unusual continuum of H-related modes puzzlingly terminating at 850 cm$^{-1}$%
, and a coalescence of higher frequency modes when the coverage approaches
one monolayer, where the anomalous continuum also appears. Balden {\it et al.%
}\cite{balden}{\it \ }noted this, and{\it \ }suggested a 2D hydrogen fluid,
as the explanation. This is however puzzling, in view of the apparent
independence of all phenomenology from either temperature, or isotopic
substitution of D in place of H. This aspect of the phenomenon therefore
remains open to further investigations.

\section{Surface Insulator-Metal transition}

So far, we have seen how a 2D metallic character may become unstable on
surfaces. In this section, I want to present cases where, somewhat to the
contrary, the metallic character may be the one eventually preferred at the
surface of a nonmetal, or of a poor metal. There is in the literature no
general discussion, or basic theory, of the SIM transition, and I am not
going to produce one here. I shall simply illustrate the notion by
discussing a few cases where I believe a SIM is taking place. This
phenomenon, in any event, is far less established than that of SMI. The most
urgent effort in this area should be experimental, for my examples are, as I
said, vastly (although not totally\cite{modesti}) derived from calculations.%
\cite{ga93}\cite{stumpf95}\cite{takeuchi94}.

\subsection{Alpha-gallium(001)}

$\alpha -$Ga is an unusual metal in many ways. It consists of covalently
bonded Ga$_2$ molecules or dimers, which have enough overlap to metallize,
along (buckled) planes. Parallel to these planes, the metallic conductivity
is reasonably high. Between the planes, however, there is mostly covalent Ga$%
_2$ bonding, and the conductivity is poorer, semi-metallic. In the bulk Ga
phase diagram, semi-metallic $\alpha -$Ga has the lowest energy under
ordinary conditions. However, other phases, such as Ga III, compete closely
with $\alpha -$Ga. In these phases, molecular bonds and covalency are
absent, and metallicity is unmitigated, as in Al. At sufficiently high
pressure and temperature, the fully metallic phases eventually prevail\cite
{Ga}.

What should the stable structure of a close-packed surface of $\alpha -$ Ga
, such as the (001) face, look like? On the ideal $\alpha -$Ga(001) face,
the Ga$_2$ dimers stick out at an angle, giving rise to an unrealistically
high charge corrugation, and, in correspondence, an unacceptably high
surface energy. Alternatively, the surface can be smoothed by breaking
surface dimers, and removing one Ga atom from each of them. However, the
surface energy is again unexpectedly high. As it turns out, the reason now
is that{\it \ }dangling bond surface states{\it \ }are formed\cite
{bernasconi95}. Eventually, calculations suggest that both the unrealistic
corrugation, and the costly dangling bonds, can be eliminated, and a
substantially lower energy state reached, by a simple kind of ''1$\times $%
1'' reconstruction. Here, remarkably, the two outermost surface layers have
become Ga III-like (with an in-plane expansion, as required by epitaxy on $%
\alpha -$ Ga), {\it and fully metallic.}\cite{ga93}{\it.}

While it can be argued that an atomically thin metallic overlayer will quite
generally contribute to lower the surface exchange-correlation energy of a
bulk nonmetal\cite{bernasconi93}, it is also clear that the surface
metallization which is predicted to take place on $\alpha -$Ga(001) would
not have had a chance, had the dangling bond surface states not been there
on the ideal surface, in the first place! In this sense, this SIM transition
has a point in common with the SMI's discussed in the previous section. Both
of them take place, in order to eliminate half-filled surface states.

As this might be quite an interesting SIM test case, I would like to take
this chance to remind the experimental community that, with the notable
exception of very careful STM studies\cite{zuger}\cite{zuger1}, there is as
yet no experimental study addressing specifically either structure, both
atomic and electronic, or excitations, of $\alpha -$Ga(001).

\subsection{ Be(0001)}

Hcp beryllium is another interesting semimetal, which has in common with $%
\alpha -$ Ga a clear pseudogap in the bulk electron density of states,
constituting a signature of covalent bonding. Unlike $\alpha -$ Ga, however
, it is not so easy here to connect bands with bonds, and in fact no
pairwise bulk covalent bond have been identified. Boron is another element
where the same ''problem'' arises. There, three-center bonds have been
discussed, as the way to understand covalency in real space.\cite{emin}, and
maybe three-center bonds should be invoked in Be as well. As discussed by
Stumpf at this meeting\cite{stumpf95}, in the case of Be, the covalent
semimetallic state is connected with a shorter c/a ratio, meaning that
covalent bonds for an atom in a (0001) plane have to do with the six
neighbors, three in the plane below and three in the plane above. As a
purely heuristic device, we could therefore think of these as six ''hole
bonds''. What this means is that with only two electrons, or six holes, in
the (2s,2p) shell, each Be atom could covalently share a hole with each of
its out-of-plane first neighbors\cite{hole}.

Let us now come to the Be surfaces. Stumpf has described, quite
convincingly, how at the (0001) surface, Be gives up its covalency, locally
recovering a fully metallic behavior. This is another example of SIM,
somewhat similar to that just discussed for $\alpha -$Ga(001). About the
physical reasons which drive this SIM, we can at best speculate, in the lack
of a good chemical understanding of even the bulk Be bonding.

At the speculative level, I find the hole bonding picture rather useful in
this connection. In particular, the reason for the SIM at Be(0001)can be
seen as due to the necessity to eliminate the three ''hole dangling bonds''
which the covalent structure would retain per each surface atom. Unlike $%
\alpha -$ Ga, where a major reconstruction is necessary to eliminate the
surface states, here the close-packed structure of Be makes the transition
possible by a simple outwards relaxation, a kind of displacive SIM.

Based on this picture, I can further speculate about the possible role of
temperature in this system. As T grows, lattice entropy should favor a {\it %
propagation} of the more metallic state deeper into the bulk, while
electronic forces which oppose that should act as a limiting factor. As a
result, the fully metallic state could remain confined to the first layer,
or to a small set of layers (a sort of {\it incomplete wetting} of
semimetallic Be by a fully metallic phase), or it could instead move into
the bulk, and trigger other transformations as well. It would be interesting
to pursue experimentally the question of which of the various possible
scenarios is realized.

\subsection{ Ge(111) at high temperature}

Another example of fine balance between bulk covalency and metallicity is
offered by the group IV elements Si, Ge, Sn. Semiconducting Si and Ge, in
particular, turn into good metals above their melting point. From the point
of view of electronic structure, metallization has to do with the fact the
fourfold tetrahedral atom coordination is lost, in the liquid state.

What should we expect of a semiconductor surface at high temperature? At the
experimental level, we know that surface disorder begins to appear, even if
only at the level of adatoms, at deconstruction. For instance on Si (111), 7%
$\times $7$\rightarrow $ 1$\times $1 deconstruction takes place 70\% of the
melting temperature $T_m$. On Ge(111), the c(2$\times $8)$\rightarrow $ 1$%
\times $1 deconstruction occurs at even lower temperature, around 40\% of $%
T_m$. In both cases, it can be argued that adatoms become mobile, and
diffuse on the surface. Therefore, a germ of disorder and atom diffusivity
is already present at relatively low temperatures. This might suggest a
gradual and continuous {\it surface melting} as $T_m$ is approached, a route
which is common to most crystal surfaces. This would imply the growth, in
full equilibrium, of a liquid metallic film on the semiconducting bulk, with
thickness diverging as $T\rightarrow $ $T_m$. On the other hand, we know
from phenomenology that {\it dispersion forces,} surprisingly important in
the surface melting problem, are going to make such a continuous melting
impossible on a semiconductor surface. In this case, due to a ''negative
Hamaker constant'' \cite{chen}, nucleation of a metallic film is welcome,
but its thickness is confined to remain microscopically small. This
behavior, when realized, goes under the name of ''incomplete surface
melting''.

Ge(111) is perhaps the best studied case of semiconductor surface at high
temperature. Although there still is some controversy\cite{meli}(once again
centered around He scattering data!), several groups have reported
incomplete melting transition of this surface near 1050 K\cite{vandergon}%
\cite{fadley}. Recent {\it ab inito} simulations also favor incomplete
melting, and predict that it should go hand in hand with surface
metallization\cite{takeuchi94}, for which there is now clear EELS evidence%
\cite{modesti}. Generalizing (which may not be fully authorized!), I would
venture to say that this example shows that SIM transitions are to be
expected on semiconductor surfaces close enough to the melting point.

Unlike the T = 0 cases discussed above, the present case of a
high-temperature SIM transition is not apparently related to eliminating
surface states of the semiconductor. On the contrary, we do predict a very
metallic density of states in the first bilayer, {\it after }the SIM has
taken place. From the point of view of local symmetry, an atom in the first
bilayer should fluctuate rapidly between a diamond, and an hexagonal diamond
configuration\cite{takeuchi94}. Photoemission from these first-bilayer
states, if feasible, should prove extremely revealing as to the real nature
and symmetry of these atoms, and of the surface metallization phase
transition altogether.

\section{Acknowledgments}

I would like to express my gratitude to Professors E. Bertel and M. Donath,
particularly for their kind insistence at having this introductive paper
written; and to my collaborators, especially A. Fasolino, C.Z. Wang, M.
Bernasconi, G.L. Chiarotti, A, Selloni, N. Takeuchi, B.N.J. Persson, and S.
Modesti. For discussions and information, I am also indebted to P.J Toennies
H.J. Ernst, E. Hulpke, and I.K. Robinson $\left[ W(001),Mo(001)\right] $, to
B. Kohler, P. Ruggerone, M. Scheffler and H. Ibach $\left[
H/Mo(110),H/W(110)\right] $, to B.N.J Persson, T. Fauster and P. Rudolf $%
\left[ K/Cu(111)\right] $, to U. D\"urig and O. Z\"uger $\left[ \alpha
-Ga\right] $, to R. Stumpf $\left[ Be\right] $, and to P.J. Toennies and G.
Lange, and J.W.M. Frenken $\left[ Ge(111)\right] $. Work at SISSA which is
related to this review was partly supported through the Italian Research
Council (CNR), Progetto SUPALTEMP, the Istituto Nazionale Fisica della
Materia (INFM), and the EEC Human Capital and Mobility scheme under
contracts No. ERBCHBGCT920180, No. ERBCHRXCT920062, and No. ERBCHRXCT930342.

\begin{figure}
\caption{Mean-field dynamical structure factor of a one-dimensional WCDW . The
acoustical phonon, of initially linear dispersion, is coupled to the
electron CDW . The resulting coupled modes are the phonon $\omega _{o\text{ }%
}$($q\leq k_F$), the phason $\omega _{-\text{ }}$($k_F\leq q\leq 2k_F$),
and the ampliton $\omega _{+\text{ }}$($q\geq 2k_F$). For visual
illustration of spectral weight, thick lines represent a ''strong''
mode. (after Giuliani and Tosatti \protect\cite{giuliani2}) Note the
resemblance with the H/W(110), and H/Mo(110) surface modes of Hulpke
and L\"udecke \protect\cite{hulpke95} \protect\cite{hulpkeH}.}
\label{fig:one}
\end{figure}


\begin{thebibliography}{999}
\bibitem{overhauser2}  A.W. Overhauser, Phys. Rev. {\bf 167} (1968) 691;
{\it Adv. Phys.} {\bf 27} (1978) 343.

\bibitem{lomer}  W. M. Lomer, {\it Proc. Phys. Soc. }{\bf 80} (1962) 489.

\bibitem{overhauser1}  A.W. Overhauser, Phys. Rev.{\bf \ 128} (1962) 1437.

\bibitem{kohn68}  W. Kohn, in {\it Many Body Physics}, ed. C. DeWitt and R.
Balian (Gordon and Breach, New York, 1968) p. 353.

\bibitem{halperin}  B.I.Halperin and T.M.Rice, {\it Solid State Physics}%
{\bf \ 21} (1968) 115.

\bibitem{salem}  H.C. Longuet-Higgins and L. Salem, {\it Proc. Roy. Soc.}%
{\bf \ A251} (1958) 172.

\bibitem{englman}  R. Englman, {\it The Jahn-Teller Effect in Molecules and
Crystals} (Wiley, New York, 1972).

\bibitem{peierls}  R.E. Peierls, {\it Quantum Theory of Solids} (Clarendon,
Oxford, 1955) Ch. 5.

\bibitem{1974}  E. Tosatti and P. W. Anderson,{\it \ Sol. State Comm. }{\bf %
14 }(1974) 713.

\bibitem{kyoto}  E. Tosatti and P. W. Anderson, {\it Proceedings of the 2nd
Int. Conf. on Solid Surfaces, Kyoto 1974, Jap. J. Appl. Phys., Suppl.}{\bf %
\ 2}{\it, }{\bf Pt. 2} (1974) 381.

\bibitem{festkorper}  E .Tosatti, {\it Festk\"orperprobleme }{\bf 15}
(1975) 113.

\bibitem{karpacz}  E. Tosatti, in {\it Modern Trends in the Theory of
Condensed Matter}, ed. A. Pekalski and J.Przystawa (Springer-Verlag,
Berlin, 1980) p. 501.

\bibitem{inglesfield79}  J. E. Inglesfield, {\it J. Phys. C} {\bf 12} (1979)
149.

\bibitem{kittel}  see, e.g., C. Kittel, {\it Introduction to Solid State
Physics,} (Wiley, New York, 1971) Ch. 17.

\bibitem{bruce}  see, e.g., A. D. Bruce and R. A. Cowley, {\it Structural
Phase Transitions }(Taylor and Francis, London, 1981).

\bibitem{pandey}  K.C. Pandey, {\it Phys. Rev. Lett.}{\bf \ 47} (1981) 1913;
{\bf 49 }(1982) 223.

\bibitem{C(111)}  S. Iarlori, G.Galli, F. Gygi, M. Parrinello, and E.
Tosatti, {\it Phys. Rev. Lett. }{\bf 69} (1992) 2947.

\bibitem{Si2x1}  F.J.Himpsel, P.M.Marcus, R.M.Tromp, I.P. Batra, M.R. Cook,
F. Jona, and H. Liu, {\it Phys. Rev.}{\bf B30} (1984) 2257.

\bibitem{takeuchi1}  N. Takeuchi, A. Selloni and E. Tosatti, {\it Phys. Rev.}%
{\bf B44} (1991) 13611.

\bibitem{Sn2x1}  Z.Y.Lu, G.L. Chiarotti, S. Scandolo, and E. Tosatti, to be
published.

\bibitem{DAS}  K. Takayanagi and Y. Tanishiro, {\it Phys. Rev. }{\bf B34}
(1986) 1034.

\bibitem{c(2x8)}  D. J. Chadi and C. Chiang, {\it Phys. Rev. }{\bf B23}
(1981) 1843.

\bibitem{ssh}  see., e.g., W.P.Su, J.R. Schrieffer and A. J . Heeger, {\it %
Phys. Rev. }{\bf B22}{\it \ }(1980) 2099.

\bibitem{anderson}  P. W. Anderson, {\it Phys. Rev. Lett.}{\bf \ 34} (1975)
953.

\bibitem{anc1}  F. Ancilotto, A. Selloni and E. Tosatti, {\it Phys. Rev. }%
{\bf B43} (RC){\bf \ }(1991) 5180.

\bibitem{biarritz}  E. Tosatti, in {\it Highlights of Condensed Matter
Physics and Future Prospects, }ed. L. Esaki (Plenum, New york, 1991) p. 631.

\bibitem{anc2}  F. Ancilotto, A. Selloni and E. Tosatti, {\it Phys. Rev. }%
{\bf B43(}RC){\bf \ }(1991) 14726.

\bibitem{dimer}  R.M.Tromp, R.G.Smeenk, F.W. Saris, and D. J. Chadi, {\it %
Surf. Sci.}{\bf \ 133} (1983) 137.

\bibitem{dimer1}  J.C. Fernandez, W.S.Yang, H.D. Shih. F. Jona, D.W Jepsen,
and P.M.Marcus, {\it J. Phys. C}{\bf 14 }(1981) L55.

\bibitem{papa}  see, {\it e.g.,}D.A. Papaconstantopoulos, {\it The Band
Structure of Elemental Solids,} (Plenum, New York, 1986).

\bibitem{harrison}  see, e.g., W.A.Harrison, {\it Electronic Structure and
the Properties of Solids}, (Freeman, S. Francisco, 1980).

\bibitem{weng}  S.L.Weng, T. Gustafsson, and E.W.Plummer, {\it Phys. Rev.
Lett. }{\bf 39} (1977) 822.

\bibitem{louie}  S. L.Louie, K.M.Ho, J.R.Chelikowsky, and M.L. Cohen, {\it %
Phys. Rev.} {\bf B15 }(1977) 5627.

\bibitem{felter}  T.E.Felter, R.A.Barker, and P.J. Estrup, {\it Phys. Rev.
Lett. }{\bf 38} (1977) 1138.

\bibitem{debe}  M.K.Debe and D. A. King, {\it J. Phys. C} {\bf 10} (1977)
L303; {\it Phys. Rev. Lett. }{\bf 39} (1977) 708.

\bibitem{1978}  E. Tosatti, Sol. State Comm. {\bf 25 }(1978) 637.

\bibitem{inglesfield78}  J.E. Inglesfield, {\it J. Phys. C} {\bf 11 }(1978)
L69.

\bibitem{hulpke95}  E. Hulpke, this volume.

\bibitem{freeman}  H. Krakauer, M. Posternak and A.J.Freeman, {\it Phys.
Rev. Lett.}{\bf 43 (}1979) 1885 .

\bibitem{ho}  X.W.Wang, C.T. Chan, K.M.Ho and W. Weber, {\it Phys. Rev.
Lett.}{\bf 60 }(1988) 2066.

\bibitem{krakauer}  D. Singh and H. Krakauer, {\it \ Phys. Rev.} {\bf B37}
(1988) 3999.

\bibitem{1980}  A. Fasolino, G. Santoro and E. Tosatti, {\it Phys. Rev.
Lett.} {\bf 44} (1980) 1684.

\bibitem{1987}  A. Fasolino and E. Tosatti, {\it Phys. Rev.} {\bf B35}{\it \ }%
(1987) 4264.

\bibitem{ernst87}  H.J.Ernst, E. Hulpke and P.J.Toennies, {\it Phys. Rev.
Lett.} {\bf 58} (1987) 1941.

\bibitem{king83}  D.A.King, {\it Physica Scr. }{\bf T4} (1983) 34.

\bibitem{ernst92}  H.J.Ernst, E. Hulpke, and P.J.Toennies, {\it \ Phys. Rev.}%
B{\bf 46} (1992) 16081, and references therein.

\bibitem{melmed}  A.J. Melmed, S.T. Ceyer, R.T.Tung, and W.R. Graham, {\it %
Surf. Sci. }{\bf 111} (1981) 1701.

\bibitem{dewette}  E. Hulpke, M. H\"uppauff, D.-M. Smilgies, A.D. Kulkarni,
and F.W. deWette, {\it Phys. Rev. }{\bf B45 }(1992) 1820.

\bibitem{erskine}  J.P.Woods and J.L.Erskine, {\it J. Vac. Sci. Technol. }%
{\bf A4} (1986) 1414.

\bibitem{wang87}  C.Z.Wang, A. Fasolino and E. Tosatti, {\it Phys. Rev.
Lett.}{\bf 59} (1987) 1845.

\bibitem{europhys88}  C.Z.Wang, M. Parrinello, A. Fasolino and E. Tosatti,
{\it Europhys. Lett. }{\bf 6} (1988) 43.

\bibitem{prb88}  C.Z.Wang, A. Fasolino and E. Tosatti, {\it Phys. Rev. }%
{\bf B37 }(1988) 2216.

\bibitem{fas88}  C.Z.Wang, A. Fasolino and E. Tosatti, {\it Europhys. Lett. }%
{\bf 7} (1988) 263 ; {\it Surf. Sci}. {\bf 211/212} (1989) 323.

\bibitem{han}  W.H.Han, S.C. Ying and D. Sahu, {\it Phys. Rev. }{\bf B41 }%
(1990) 4403.

\bibitem{unpubl91}  A. Fasolino and E. Tosatti, unpublished; E. Tosatti,
results presented at the Cambridge Meeting, March 1991.

\bibitem{smilgies91}  E. Hulpke and D.-M. Smilgies, {\it Phys. Rev. }{\bf %
B43 }(1991) 1260.

\bibitem{hildner}  M.L.Hildner, R.S.Daley, T.E. Felter and P.J. Estrup, {\it %
J. Vac. Sci. Technol. }{\bf A9} (1991) 1604.

\bibitem{M5}  A. Fasolino, G. Santoro and E. Tosatti, {\it J. Phys. (Paris)
}{\bf 42}, {\bf C6} (1981) 846.

\bibitem{Mo88}  C.Z.Wang, A. Fasolino and E. Tosatti, {\it Phys. Rev. Lett.
}{\bf 60} (1988) 2661.

\bibitem{wang95}  C.Z.Wang, private communication (1995).

\bibitem{robinson}  D.-M. Smilgies, P.J. Eng, and I.K. Robinson, {\it Phys.
Rev. Lett. }{\bf 70} (1993) 1291.

\bibitem{roelofs}  L.D. Roelofs and S.M. Foiles, {\it Phys. Rev. }{\bf B48 }%
(1993) 11287.

\bibitem{carlsson}  A.E. Carlsson, {\it Phys. Rev. }{\bf B44 }(1991) 6590.

\bibitem{campuzano}  J.C. Campuzano, D.A. King, C. Somerton, and J.E.
Inglesfield, {\it Phys. Rev. Lett. }{\bf 45 }(1980) 1649.

\bibitem{kevan90}  K. Smith, G. Elliott, and S. Kevan, {\it Phys. Rev. }{\bf %
B42 }(1990) 5385.

\bibitem{chung}  J.W.Chung, this volume, and references therein.

\bibitem{lapuj}  J. Spr\"osser, B. Salanon, and J, Lapujoulade, {\it %
Europhys. Lett.}{\bf 16} (1991) 283.

\bibitem{cvetko}  D. Cvetko, A. Lausi, A. Morgante, F. Tommasini and K.C.
Prince, {\it Surf. Sci }{\bf 269/270 }(1991) 68.

\bibitem{vlieg}  I.K. Robinson, E. Vlieg, and K. Kern, {\it Phys. Rev.
Lett. }{\bf 63} (1989) 2578.

\bibitem{kern}  M. Krzyzowsky, P. Zeppenfeld, and K. Kern, preprint (1994).

\bibitem{bernasconi93}  see, {\it e.g., }M. Bernasconi and E. Tosatti, {\it %
Surf. Sci. Repts. }{\bf 17 }(1993) 363.

\bibitem{mazzeo}  G. Mazzeo, G. Jug, A. C. Levi, and E. Tosatti, {\it Surf.
Sci }{\bf 273 }(1992) 237.

\bibitem{dennijs}  M. Den Nijs, {\it Phys. Rev. Lett. }{\bf 66} (1991) 907.

\bibitem{heine88}  V. Heine and J.J.A. Shaw, {\it Surf. Sci }{\bf . 193 }%
(1988) 153.

\bibitem{robinson1}  I.K. Robinson, private communication (1993).

\bibitem{wang thesis}  C.Z. Wang, PhD Thesis, SISSA, Trieste, November 1987.

\bibitem{Cu(111)}  S.D. Kevan, {\it Phys. Rev. Lett. }{\bf 50 }(1983) 526.

\bibitem{shoenberg}  D. Shoenberg and D.J.Roaf,{\it \ Phil. Trans.Roy. Soc.}
{\bf 255} (1962) 85 .

\bibitem{inglesfield95}  S. Crampin, M.H.Boon, and J.E. Inglesfield, {\it %
Phys. Rev. Lett.}{\bf 73} (1994) 1015; and{\it,} this volume.

\bibitem{kevan95}  S.D. Kevan, this volume.

\bibitem{eigler}  M.F. Crommie, C.P.Lutz, and D.M.Eigler, Nature {\bf 363 }%
(1993) 524; M.F. Crommie, C.P.Lutz, D.M.Eigler and E.J.Heller, this volume.

\bibitem{einstein}  T.B. Einstein and J.R. Schrieffer, {\it Phys. Rev. }{\bf %
B7 }(1973) 3629.

\bibitem{schrieffer}  J.R.Schrieffer and P. Soven, {\it Physics Today} {\bf %
28} (1975) 24.

\bibitem{1976}  E. Tosatti, in {\it Proceedings of the 13th Int. Conf. on
the Physics of Semiconductors,} ed. F.G. Fumi (North Holland, 1976) p. 21.

\bibitem{kohn}  K.H.Lau and W. Kohn, {\it Surf. Sci }{\bf . 75} (1978) 79.

\bibitem{hirschmugl}  F.M.Hoffmann, B.N.J.Persson, W. Walter, D.A. King, C.J
Hirschmugl, and G.P. Williams, {\it Phys. Rev. Lett.} {\bf 72} (1994) 1256.

\bibitem{persson95}  B.N.J.Persson, F.M Hoffmann and G.P. Williams, this
volume.

\bibitem{persson}  B.N.J.Persson and A.I. Volokitin,{\it \ Surface Sci.}
{\bf 310} (1994) 314.

\bibitem{fauster}  N.Fischer, S. Schuppler, R. Fischer,Th. Fauster, and W.
Steinmann, {\it Phys. Rev. {\bf B47 }}(1993) 4705.

\bibitem{fauster1}  N.Fischer, S. Schuppler, Th. Fauster, and W. Steinmann,
{\it Surf. Sci }{\bf . 314 }(1994) 89.

\bibitem{K/Cu(111)}  N.Fischer, S. Schuppler, R. Fischer, Th. Fauster, and
W. Steinmann, {\it Phys. Rev.}{\bf B(RC) 43 }(1991) 14722.

\bibitem{hulpkeH}  E.Hulpke and J. L\"udecke, {\it Phys. Rev. Lett.{\bf \ 68}%
} (1992) 2846; {\it J.} {\it Electron Spectr. Rel. Phenom.}{\bf \ 64/65}
(1993) 641.

\bibitem{carneiro}  K. Carneiro, G. Shirane, S.A.Werner, and S. Kaiser,
{\it Phys. Rev. }B {\bf 13} (1976) 4258.

\bibitem{giuliani1}  G.Giuliani and E. Tosatti, {\it Il Nuovo Cimento}{\bf \
47} B (1978) 135.

\bibitem{giuliani2}  G.Giuliani and E. Tosatti, in {\it \ Quasi
One-Dimensional Conductors I}, ed S. Barisic, A. Bjelis, J.R. Cooper, and
B. Leontic (Springer-Verlag, Berlin,1979) p. 191.

\bibitem{kevan89}  R.H. Gaylord, K.H.Jeon, and S.D.Kevan, {\it Phys. Rev.
Lett. }{\bf 62} (1989) 2036.

\bibitem{kohler}  B. Kohler, P. Ruggerone, S. Wilke, and M. Scheffler, {\it %
Phys. Rev. Lett. {\bf 74 }}(1995) 1387{\it ; }P. Ruggerone, B. Kohler, S.
Wilke, and M Scheffler, this volume.

\bibitem{balden}  M. Balden, S. Lehwald, H. Ibach, and D.L. Mills, {\it %
Phys. Rev. Lett. }{\bf 73} (1994) 855.

\bibitem{modesti}  S. Modesti, V.R. Dhanak, M. Sancrotti, A. Santoni, B.N.J.
Persson, and E. Tosatti, {\it Phys. Rev. Lett. }{\bf 73 }(1994) 1951.

\bibitem{ga93}  M. Bernasconi, G.L.Chiarotti, and E. Tosatti, {\it Phys.
Rev. Lett. }{\bf 70 }(1993) 3295.

\bibitem{stumpf95}  R. Stumpf, J.B.Hannon, and E.W.Plummer, this volume,
and references therein.

\bibitem{takeuchi94}  N.Takeuchi, A. Selloni, and E. Tosatti, {\it Phys.
Rev. Lett. }{\bf 72} (1994) 2227.

\bibitem{Ga}  M. Bernasconi, G.L.Chiarotti, and E. Tosatti, {\it Phys. Rev.}%
{\bf B, }to appear, and references therein.

\bibitem{bernasconi95}  M. Bernasconi, G.L.Chiarotti, and E. Tosatti, {\it %
Phys. Rev.}{\bf B, }to appear.

\bibitem{zuger}  O. Z\"uger and U.D\"urig, {\it Phys. Rev.}{\bf B46 }(1992)
7319.

\bibitem{zuger1}  O. Z\"uger, PhD Thesis, ETH Zurich No. 9658 (1992).

\bibitem{emin}  D. Emin, {\it Physics Today} {\bf 40 }(1987) 55.

\bibitem{hole}  If unoccupied atomic shells like 3s, 3p, etc. were
sufficiently high in energy, a hole representation within the (2s, 2p) shell
would be rigorously equivalent to the electron representation. Hole charge
density maps, obtained by subtracting the actual electron density from a
fictitious ''parent'' density obtained by completely filling the (2s, 2p)
shell with 8 electrons/atom, could reveal possible pairwise hole bonds,
invisible in the electron maps. However, due to the closeness in energy of
the higher shells, it is not obvious that in practice a hole representation
can be given a quantitative meaning in Be. It is therefore used here merely
as a heuristic device.

\bibitem{chen}  X.J. Chen, A.C. Levi, and E. Tosatti, {\it Il Nuovo Cimento }%
{\bf D13 }(1992) 919.

\bibitem{meli}  C.A.Meli, E.F. Greene, G. Lange, and P.J.Toennies, {\it %
Phys. Rev. Lett. {\bf 74 }}(1995) 2054.

\bibitem{vandergon}  A.W.Denier van der Gon {\it et al., Surf. Sci. }{\bf %
241 }(1991) 335.

\bibitem{fadley}  T.T. Tran {\it et al.}, {\it Surf. Sci. }{\bf 281 }(1993)
270.
\end{thebibliography}
\end{document}